\documentclass{article}

\usepackage[final,main]{neurips_2025}
\usepackage{hyperref}
\usepackage[utf8]{inputenc}
\usepackage[T1]{fontenc}
\usepackage{bm}
\usepackage{xspace}
\usepackage{amsmath}
\usepackage{amsfonts}
\usepackage[ruled,vlined]{algorithm2e}
\usepackage{physics}
\usepackage{graphicx}
\usepackage[caption=false]{subfig}
\usepackage{booktabs}
\usepackage{xcolor}
\usepackage[capitalise]{cleveref} % must be after amsmath and hyperref
\usepackage{verbatim}
\newcommand{\columbia}{Department of Astronomy, Columbia University, New York, NY, 10027}

\newcommand{\tail}{\textsc{Tailed-Uniform}\xspace}
\newcommand{\uniform}{\textsc{Uniform}\xspace}
\hypersetup{
    colorlinks=true,
    linkcolor=black,
    citecolor=purple,
    urlcolor=blue,
    }

\graphicspath{{./}{figures/}}

\begin{document}

\title{Learning at the Edge: Tailed-Uniform Sampling for Robust Simulation-Based Inference}
\author{%
  Chaipat Tirapongprasert$^{1*}$ \quad
  Matthew Ho$^{1}$ \\
  $^{1}$\columbia
}

\maketitle

\begin{abstract}
We introduce the \tail proposal distribution for generating training simulations in simulation-based inference. Instead of sampling parameters uniformly within bounded regions, we extend the distribution beyond prior boundaries with smooth Gaussian tails. This eliminates sharp transitions that cause neural posterior estimators to fail when the posterior distribution intersects or extends beyond the prior bounds. We show these benefits on a toy problem and cosmological parameter inference from the matter power spectrum.
Such an advantage grows in high dimensions, where boundaries dominate parameter space volume.
\end{abstract}

\section{Introduction} \label{sec:intro}

% Today, researchers, among them astronomers, face a growing challenge, as many systems of interest (e.g., exoplanets, black hole binaries, or galaxy clusters) are not amenable to direct probing \citep{feigelson2012,dodelson2020}. Thus emerges a new class of problems known as ``inverse problems," in which the key goal is to identify the model parameters that generated some observed data. The key challenge is that almost all mathematical models of real-world phenomena are complex in that they demand high-dimensional parameter space. This makes it prohibitively expensive to compute the probability density (i.e., likelihood) for some observation.

Modern astrophysics increasingly relies on complex forward simulations to model non-linear phenomena, ranging from the formation of large-scale structure \citep{pakmor2023millenniumtng} to binary black hole evolution \citep{siwek2023orbital}. As the fidelity of these physical models improves, so does the dimensionality of their parameter space, creating a significant bottleneck for inverse problems where the goal is to recover governing parameters from observations. Traditionally, parameter estimation relied on Markov Chain Monte Carlo \citep[MCMC;][]{metropolis1953,hastings1970}, which necessitates the computation of an explicit, closed-form likelihood. As the complexity and dimensionality of our models grow, this is typically not possible or too costly \citep{pretorius2005}.

Simulation-based inference \citep[SBI;][]{cranmer2020} has emerged as the standard framework for tackling intractable likelihoods and is now a cornerstone methodology in physics, spanning cosmology, gravitational-wave astronomy, and particle physics. Most SBI methods rely on neural density estimation \citep{papamakarios2018,greenberg2019,lueckmann2017}; we focus on Neural Posterior Estimation (NPE), which trains on simulated $(\boldsymbol\theta, \mathbf{x})$ pairs to directly approximate the posterior and amortizes the simulation cost so that one trained network can infer posteriors for any new observation.

However, the fidelity of the amortized posterior depends on having a diverse and representative training set, which begins with the choice of proposal distribution from which simulation parameters are drawn.
Often, simulators preparing datasets for large-scale public campaigns, such as the CAMELS \citep{villaescusa-navarro2021}, choose to simulate at a uniform distribution of points in parameter space, typically via a Latin Hypercube bounded within a specific region of interest \citep{mckay2000}. While efficient for volume coverage, this strategy introduces a sharp discontinuity in the probability density at the prior boundaries. Neural density estimators, driven by stochastic gradient descent, struggle to model this artificial step-function \citep{cornish2020relaxing}. Consequently, the estimators often fail to capture the posterior accurately near the edges of the parameter space. This boundary pathology is exacerbated in high-dimensional settings by the curse of dimensionality \citep{bellman2003}, where the volume of the parameter space is increasingly dominated by the boundary shell rather than the interior.

To address these problems, we present \tail, a hybrid proposal distribution that combines \uniform cores with smooth Gaussian tails beyond the prior boundaries. Neural networks trained with \tail achieve superior posterior quality near boundaries while maintaining consistent performance across the full parameter space, with advantages growing in higher dimensions. We validate \tail on a Gaussian linear benchmark and on cosmological parameter inference from the matter power spectrum. All code is publicly available on Github.\footnote{\url{https://github.com/chaipattira/tailed-uniform-sbi.}}

\section{Methods} \label{sec:methods}
We define a simulator as a function $\mathcal{M}$ that maps parameters $\boldsymbol{\theta} \in \Theta \subseteq \mathbb{R}^d$ to observables $\mathbf{x} \in \mathcal{X} \subseteq \mathbb{R}^D$ \citep{cranmer2020}. Training data consists of pairs $\{(\boldsymbol{\theta}_i, \mathbf{x}_i)\}$ generated by sampling from a proposal distribution and running the forward model.
\begin{comment}
\begin{figure}
    \centering
\includegraphics[width=0.75\linewidth]{figures/uniform_vs_tailed_uniform.pdf}
    \caption{The standard \uniform distribution (blue) has constant probability density between -1 and 1 with zero probability outside this range. The \tail distribution (magenta) maintains a uniform density in the same central region $[-1, 1]$ but extends beyond these boundaries with Gaussian tails characterized by standard deviation $\sigma = 0.2$.}
    \label{fig:distribution}
    \vspace{0.5em}
\end{figure}
\end{comment}
We construct a neural architecture $q_w(\boldsymbol\theta\mid\mathbf x)$ that approximates the posterior $\mathcal{P}(\boldsymbol\theta\mid\mathbf x)$ by minimizing the negative log-probability loss $\mathcal{L}_{\text{NPE}} = -\mathbb{E}_{\mathcal{D}_{\text{train}}} \log \left[\frac{\mathcal{P} (\boldsymbol\theta)}{\tilde{\mathcal{P}}(\boldsymbol\theta)} q_w(\boldsymbol\theta|\mathbf{x})\right]$, where the expectation is over training pairs $\{(\boldsymbol\theta_i, \mathbf{x}_i)\}_{i=1}^N$ \citep{papamakarios2018}.

\textbf{Priors} \;\; It is crucial to distinguish
between proposal and assumed priors. The proposal prior $\tilde{\mathcal{P}}(\boldsymbol\theta)$ is the empirical distribution of training parameters, while the assumed prior $\mathcal{P}(\boldsymbol\theta)$ encodes scientific beliefs or previous
empirical knowledge about plausible parameter values \citep{cranmer2020}. Note that when we sample training simulations from a \uniform distribution $\tilde{\mathcal{P}} = \mathcal{U}([\theta_{\min}, \theta_{\max}]^d)$, the density drops sharply to zero at the prior boundaries. Because the network can only learn densities where training data exist, posteriors near boundaries are systematically unreliable.

\subsection{\tail Proposal}
By assigning weights beyond the primary region of interest, our proposal offers a smooth, continuously differentiable density across $\mathbb{R}^d$, alleviating the sharp discontinuities at the boundary. In a one-dimensional ($d = 1$) case, we define the probability density function as
\begin{equation}
\tilde{\mathcal{P}}_{\text{\tail}}(x; a, b, \sigma) = \begin{cases}
A \cdot \mathcal N(a, \sigma^2), & x \leq a \\
B \cdot \mathcal U(a, b), & x \in [a, b] \\
A \cdot \mathcal N(b, \sigma^2), & x \geq b
\end{cases}
\end{equation}
where $a$ and $b$ define the boundaries of the \uniform core region, and $\sigma$ controls the width of the Gaussian tails.
We also impose continuity at the boundary to guarantee that our distribution is well-defined, which in turn establishes the values of the normalization constants as
$A = \frac{\sigma\sqrt{2\pi}}{\sigma\sqrt{2\pi} + (b-a)} \text{ and } B = \frac{b-a}{\sigma\sqrt{2\pi} + (b-a)}$.
To generalize to a multivariate parameter space $\boldsymbol{\theta} \in \mathbb{R}^d$, we write it as a product of independent univariate \tail distributions
\begin{equation}
\tilde{\mathcal{P}}_{\text{\tail}}(\boldsymbol{\theta}; \mathbf{a}, \mathbf{b}, \boldsymbol{\sigma}) = \prod_{i=1}^{d}\tilde{\mathcal{P}}_{\text{\tail}}(\theta_i; a_i, b_i, \sigma_i),
\end{equation}
where $\mathbf{a} = (a_1, \ldots, a_d)$ and $\mathbf{b} = (b_1, \ldots, b_d)$ define the hypercube boundaries, and $\boldsymbol{\sigma} = (\sigma_1, \ldots, \sigma_d)$ controls the tail-widths in each dimension.
Smaller values of $\sigma_i$ concentrate more samples within the target region $[a_i, b_i]$ but provide less smoothing at the boundaries. In practice, we recommend setting $\sigma_i$ as about $10-30$ percent of the box width.

\section{Toy Problem} \label{sec:toy}
We first demonstrate \tail on a toy problem where the true posterior is analytically tractable.
We use the Gaussian Linear task from the simulation-based inference benchmark \citep[\texttt{sbibm};][]{lueckmann2021} to construct a 2D Gaussian simulator $\mathcal{M}(\boldsymbol{\theta}) = \boldsymbol{\theta}$ with a standard Gaussian prior $\mathcal{N}(\mathbf{0}, \mathbf{I}_2)$ on $\boldsymbol{\theta} \in [-1,1]^2$, giving a Gaussian likelihood $\mathcal P(\mathbf{x}\mid\boldsymbol{\theta}) = \mathcal{N}(\mathbf{x}; \boldsymbol{\theta}, \mathbf{I}_2)$.

\textbf{Training} \;\; We generate $N = 6{,}000$ pairs $\{(\boldsymbol{\theta}_i, \mathbf{x}_i)\}$ from two  NPE models trained with distinct
strategies (proposals): a \textit{\uniform} baseline $\mathcal{U}([-1,1]^2)$ and our \tail with tail width $\sigma_i = 0.1\,(b_i - a_i)$.
Built on the \texttt{LtU-ILI} framework \citep{ho2024}, both NPE models use ensembles of Masked Autoregressive Flows (MAF) \citep{papamakarios2018maf} and Masked Autoencoders (MADE) \citep{germain2015} with 16 hidden features and 5 coupling layers.
\begin{figure}
    \centering
\subfloat[\textit{\uniform}: boundary degradation (blue near edges).]{%
        \includegraphics[width=0.49\linewidth]{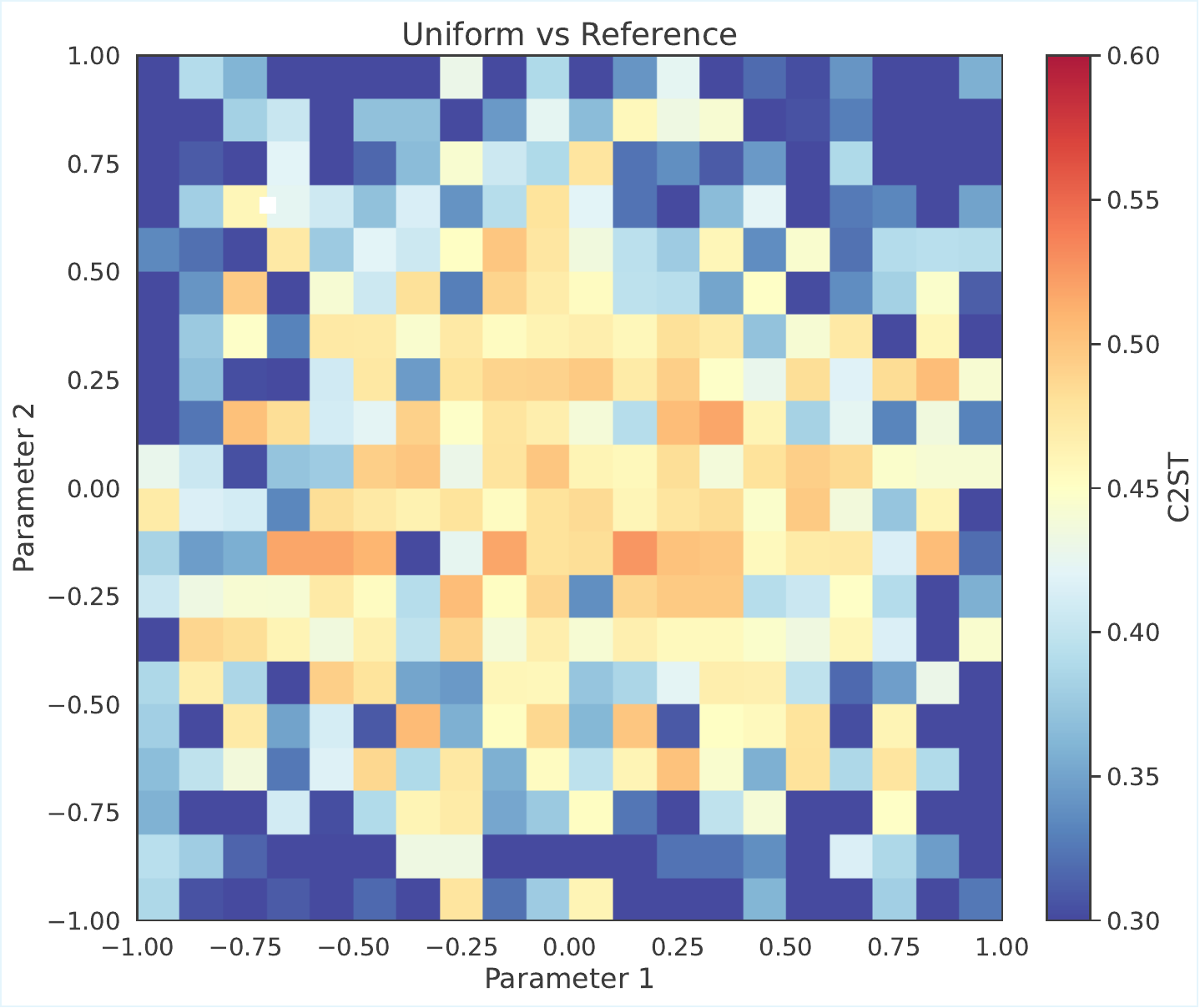}
    }\hfill
    \subfloat[\tail: improved, more consistent performance across the full space.]{%
        \includegraphics[width=0.49\linewidth]{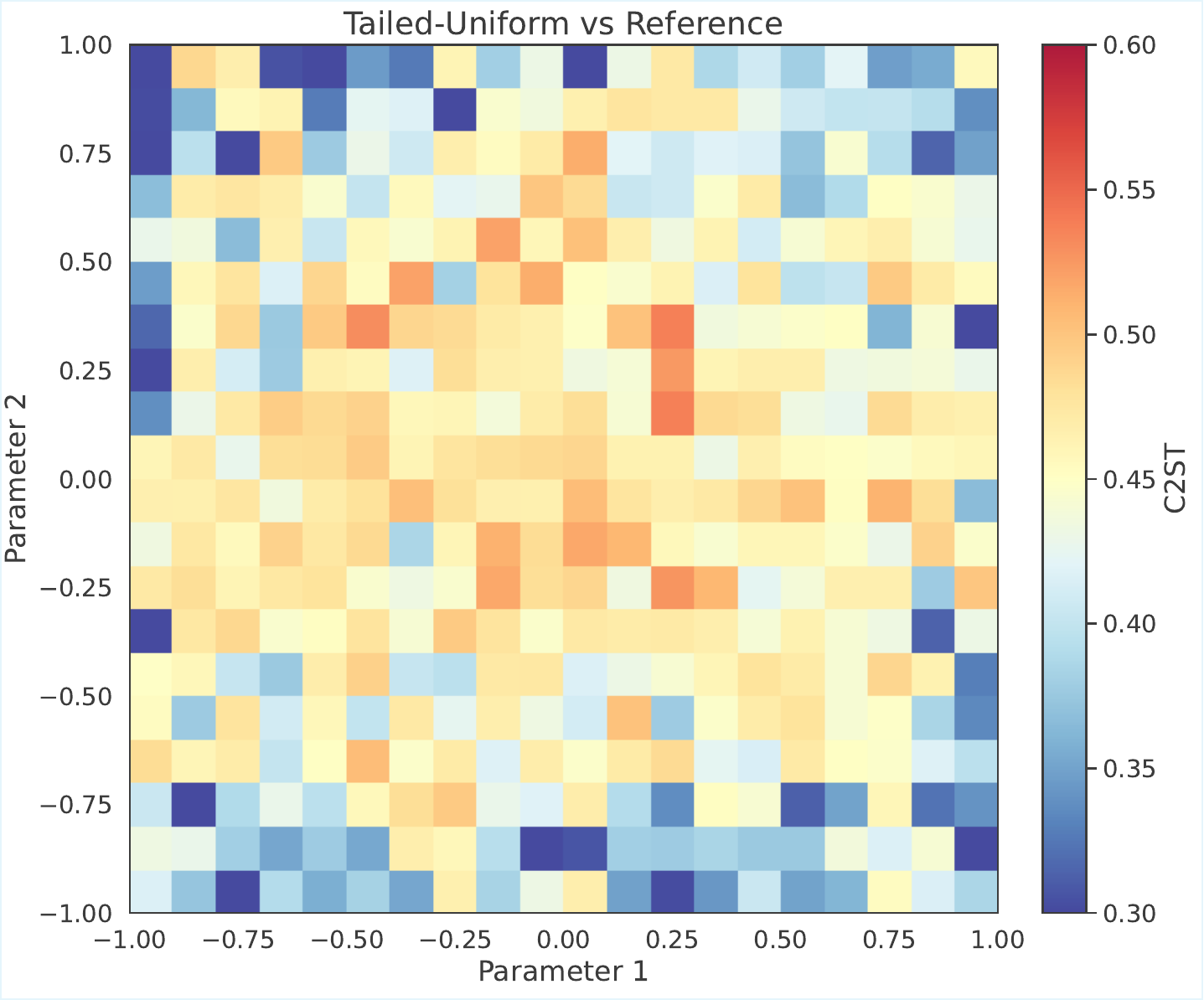}
    }
    \caption{C2ST scores across the parameter space. Blue $\ll 0.5$ = poor; orange/red $\approx 0.5$ = good.}
    \label{fig:c2st_heatmaps}
\end{figure}

\textbf{Validation} \;\; 
To measure \tail's performance across the entire parameter space, we discretize a uniform rectangular grid with $n = 20$ test locations along each dimension of $[-1, 1]^2$, yielding 400 evaluation points ranging from the center to the boundaries. For each point, we generate an observation, draw $M = 1000$ posterior samples from all three methods, and compute pairwise Classifier Two-Sample Test (C2ST) scores \citep{lopez-paz2018}. C2ST trains an independent classifier to distinguish samples from two distributions; a score of 0.5 indicates identical distributions (perfect inference). 

Figure~\ref{fig:c2st_heatmaps} shows how the C2ST degradation
pattern is rotationally invariant, depending only on radial distance from the center:
\textit{\uniform}'s performance deteriorates as it drifts away from the prior center, as evidenced by the proliferation of blue patches near the boundary. In contrast, our \tail maintains more red and orange coloration (with C2ST scores around $0.44$--$0.49$) across the parameter space.

\begin{figure}
    \centering
    \includegraphics[width=0.95\linewidth]{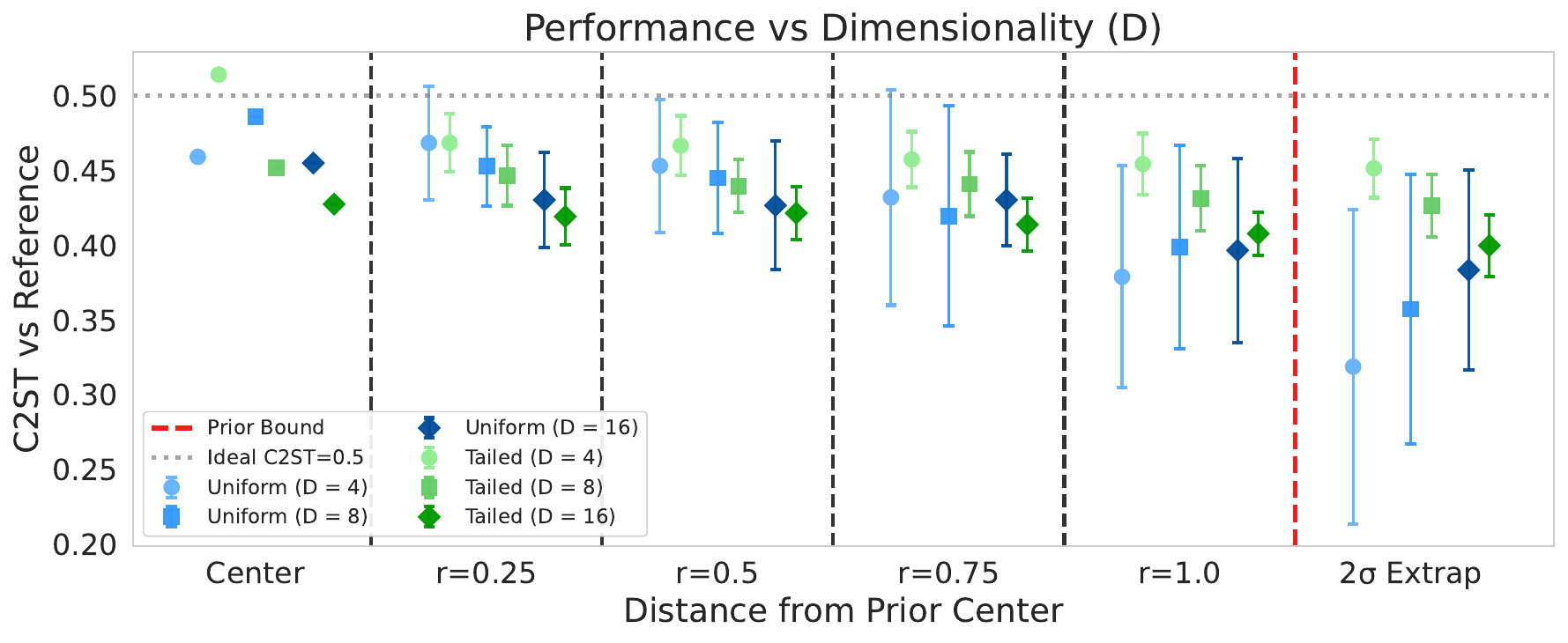}
    \caption{C2ST vs.\ distance from center across dimensions. \tail maintains superior performance over \uniform, with the gap generally widening at higher $d$, though the relative advantage varies across radial distances. Error bars: 16th--84th percentile.}
    \label{fig:dim}
\end{figure}
\textbf{Scaling to higher dimensions} \,\, By the curse of dimensionality, boundary regions of a $d$-dimensional hypercube dominate the total volume exponentially as $d$ grows, making boundary pathology increasingly severe for high-dimensional inference tasks common in physics.
Although \tail allocates a growing fraction of samples to the Gaussian-tail regions beyond $[a_i, b_i]$ (24\% at $d=2$, 86\% at $d=16$), Figure~\ref{fig:dim} shows that \tail maintains superior performance over \textit{\uniform} across all tested dimensions $d \in \{4, 8, 16\}$.
At high dimensions, boundary regions are so dominant that the marginal value of tail samples outweighs the cost of diluting interior coverage.

\section{Science Experiment} \label{sec:sci}
Following the same training and validation pipeline in Section~\ref{sec:toy}, we now apply \tail to a popular inference benchmark in cosmology: the inference of cosmological parameters $\boldsymbol{\theta} = (\Omega_m, h)$ from the matter power spectrum $\mathbf{P} \in \mathbb{R}^{64}$, where $\Omega_m$ is the matter density parameter and $h$ is the dimensionless Hubble parameter.
We generate observations using the \texttt{syren-new} emulator \citep{sui2025} with heteroskedastic cosmic variance noise \citep{dodelson2020}.
As reference posteriors, we use MCMC to collect samples under a truncated Gaussian prior centered in $\Omega_m \in [0.24, 0.40]$, $h \in [0.61, 0.73]$. To ensure a fair comparison between the NPEs, we perform independent \texttt{Optuna} hyperparameter searches \citep{akiba2019} for each proposal over flow architecture (MAF or NSF), hidden features, transform layers, batch size, and learning rate.

\textbf{Degeneracy} \,\,
The matter power spectrum is sensitive to the combination $\Omega_m h$ rather than to each parameter independently \citep{peacock1996}, creating an anti-correlation that manifests as slanted banana-shaped posterior contours. 
As a result, the information embedded in the posterior will be weak along the degenerate directions in the parameter space. To connect the performance degradation trend to the underlying physics, we propose a directional binning strategy along the degeneracy direction $d_1 = \Omega_m + h$ and the perpendicular direction $d_2 = \Omega_m - h$.

\begin{figure}
\centering
\subfloat[Degeneracy direction ($d_1 = \Omega_m + h$): \textit{\uniform} degrades at both ends; \tail is stable.]{%
    \includegraphics[width=0.49\linewidth]{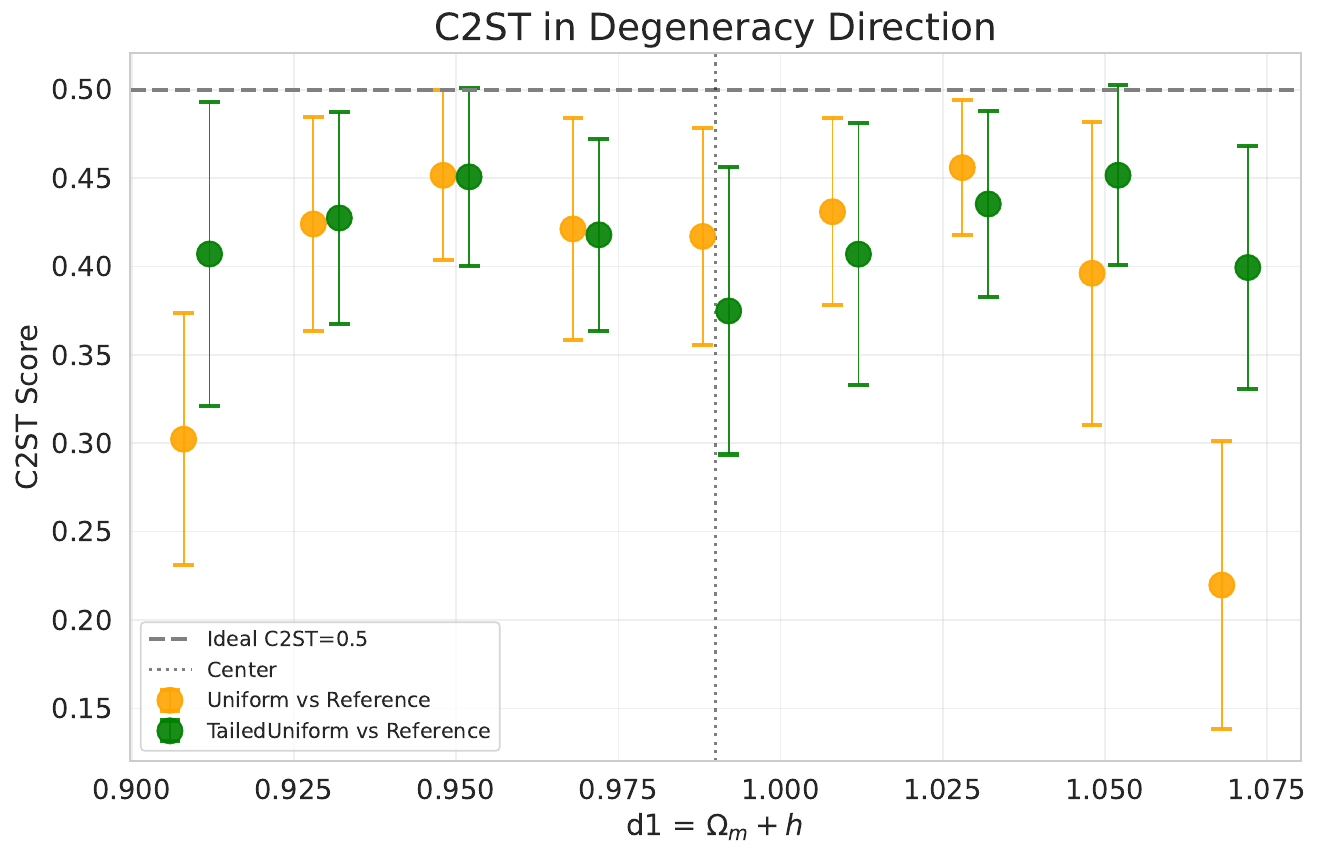}
}\hfill
\subfloat[Perpendicular direction ($d_2 = \Omega_m - h$): both methods perform comparably.]{%
    \includegraphics[width=0.49\linewidth]{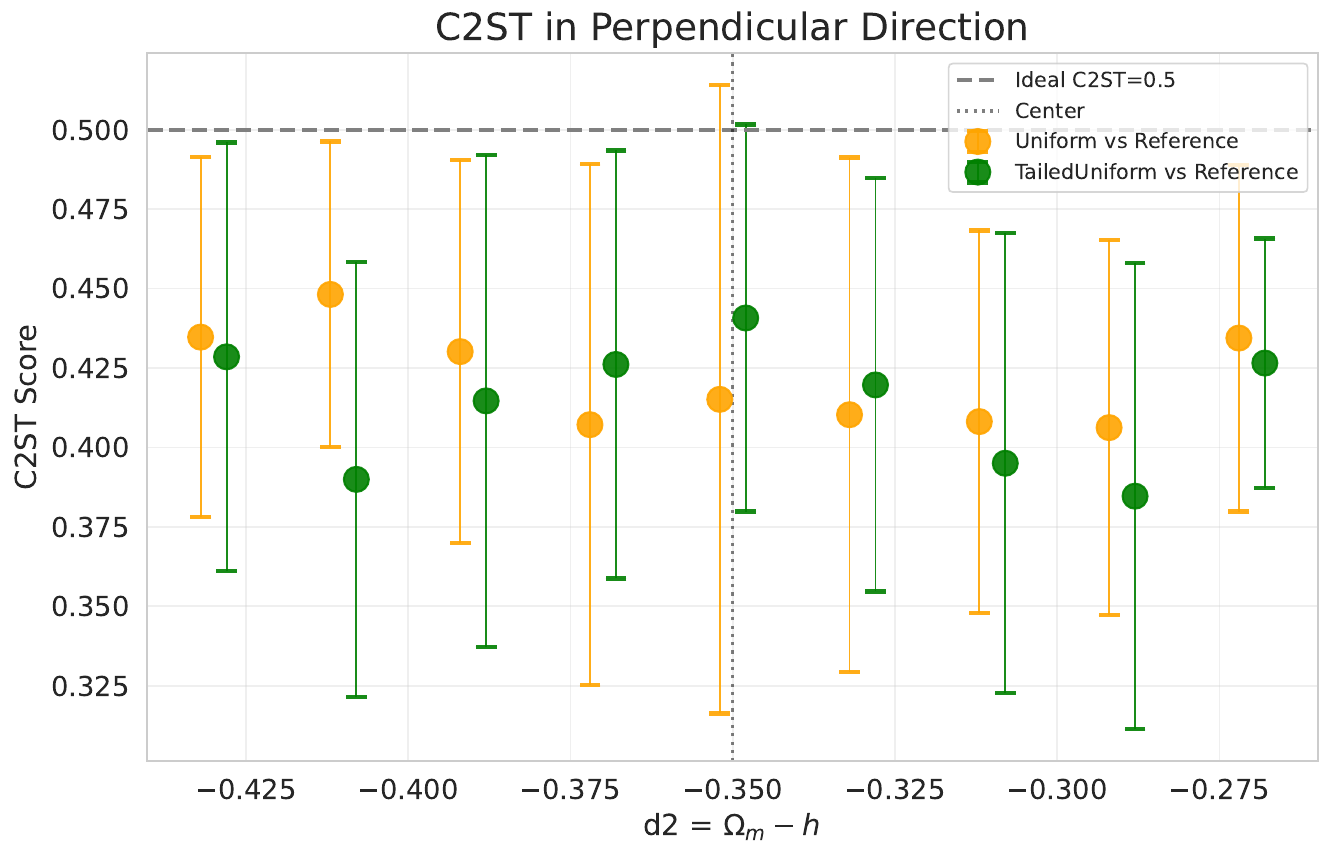}
}
\caption{C2ST vs.\ position along two orthogonal directions. Error bars: 16th--84th percentile.}
\label{fig:directional-sci}
\end{figure}

Figure~\ref{fig:directional-sci} shows that boundary pathology is anisotropic, with the most severe manifestation along degenerate directions. While both approaches perform comparably (excluding statistical fluctuations) along the perpendicular direction $d_2$, the \textit{\uniform} performance sinks at both boundaries across the degeneracy direction $d_1$.

\section{Conclusion} \label{sec:conclusion}

We proposed \tail, a hybrid proposal distribution that replaces the sharp boundary of the standard \uniform prior with smooth Gaussian tails, mitigating the structural pathology that causes NPE posteriors to degrade near the edges of parameter space.
With minimal hyperparameter tuning, \tail enables more robust inference across the full parameter space.

In the Gaussian Linear benchmark with a Gaussian (assumed) prior, \tail maintains more consistent C2ST scores throughout the domain while \uniform degrades near the boundary. The advantage also appears to widen in higher dimensions where boundary regions dominate the parameter volume. For cosmological parameter inference, \tail provides support along the degenerate direction where elongated posteriors extend toward prior edges.
Future work should explore behavior beyond $d=16$ and whether aligning the proposal tails with the posterior degeneracy direction rather than the physical parameter axes yields further gains.

\newpage
\bibliography{references}
\end{document}